\documentstyle[aps,floats,epsf]{revtex}  

\draft
\begin{document}

\twocolumn[\hsize\textwidth\columnwidth\hsize\csname
@twocolumnfalse\endcsname

\title{Electrostatic model of atomic ordering in complex perovskite
alloys}

\author{L. Bellaiche and David Vanderbilt}

\address{Department of Physics and Astronomy,\\
         Rutgers University, Piscataway, New Jersey 08855-0849, USA}

\date{June 23, 1998}

\maketitle

\begin{abstract}
We present a simple ionic model which successfully reproduces the
various types of compositional long-range order observed in 
a large class of complex
insulating perovskite alloys.  The model assumes that the driving
mechanism responsible for the ordering is simply the electrostatic
interaction between the different ionic species.  A possible new
explanation for the anomalous long-range order observed in some Pb
relaxor alloys, involving the proposed existence of a small amount
of Pb$^{+4}$ on the B sublattice, is suggested by an analysis of
the model.
\end{abstract}


\vskip2pc]

\narrowtext

Complex insulating perovskite alloys are of
great interest for their actual or potential uses mainly because
of their exceptional dielectric and piezoelectric properties
\cite{Lines}.  Examples are the so-called ``super Q'' mixed metal
perovskites, such as
Ba(Zn$_{1/3}$Ta$_{2/3})$O$_{3}$ (BZT) or
Ba(Mg$_{1/3}$Nb$_{2/3})$O$_{3}$--BaZrO$_{3}$ (BMN--BZ), which can
be used in high frequency applications because of their ultra-low
losses at microwave frequencies \cite{Akbas2,Matsumoto}.  Other
examples are the ``new'' relaxor ferroelectric alloys, such as
Pb(Mg$_{1/3}$Nb$_{2/3}$)O$_{3}$--PbTiO$_{3}$ (PMN--PT) or
Pb(Zn$_{1/3}$Nb$_{2/3}$)O$_{3}$--PbTiO$_{3}$ (PZN--PT), which
exhibit extraordinarily high values of the piezoelectric constants
\cite{Park} and thus promise to improve the resolution and
range of ultrasound and sonar listening devices \cite{Service}.

An aspect that is not well understood about these complex materials
is the compositional atomic ordering that occurs in some compounds
and not in others, and the different kinds of long-range order
(LRO) that can occur.  For example, increasing progressively the
bulk composition $x$ in
$(1-x)$~Ba(Mg$_{1/3}$Nb$_{2/3}$)O$_{3}$~+~$x$~BaZrO$_{3}$ (BMN-BZ)
leads to the following types of order for a final sintering temperature
of 1640$^\circ$C \cite{Akbas2}: the so-called ``1:2''
ordering along the [111] direction for very small values of $x$ (i.e.,
lower than 5\%); then ``1:1'' ordering along [111] for compositions
ranging from 5 to 25\%; and finally, disordering for larger values
of $x$.

The nature and strength of the compositional order may have
crucial consequences for the desired properties of the material.
For example, previous research has shown that  the microwave
loss properties of perovskite ceramics are very sensitive to the
B-site cation order with the $Q$ value improving with increasing
degree of order \cite{Matsumoto,Kawashima}. Similarly,
compositional fluctuations are believed to play a central role
in the ``relaxor'' behavior in Pb-based perovskite alloys \cite{Setter}.

Previous pioneering theoretical work \cite{Burton1,Burton2} has
been partially successful in describing the ordering in some
cases. In such models, the Coulomb interactions are included
indirectly via a cluster expansion of finite range.  While there
is some justification for such an approach \cite{Ceder}, the
ordering sequences that occur in complex materials such as BMN-BZ
did not emerge naturally, and thus the driving mechanism
responsible for the ordering has remained somewhat obscure.

Motivated to resolve these problems, we decided to investigate
a simple model that explicitly includes the long-range Coulomb
interactions between ions. In this model, the energy is taken to
be proportional to the electrostatic energy of an ideal system
of ionic charges.  We initially expected that this very simple
model might need to be augmented with corrective covalent terms
before it could provide a useful account of the different orderings
seen in real materials.  To our surprise, the minimal model by
itself is remarkably successful.  The purpose of this letter,
then, is to show that this very simple model provides a systematic
understanding of the complicated ordering behavior of complex
perovskite alloys, thereby providing strong evidence that the
Coulomb interaction between ions is the dominant factor in
determining such ordering.

Our model assumes that the minimization
of the electrostatic interactions between the different ions in
the compounds leads to the LRO seen experimentally. The interaction
energy in ABO$_{3}$ perovskite compounds is then expressed as
\begin{equation}
E = {e^2\over2} \sum_{(l\tau)\ne(l'\tau')}
  {Q_{l\tau}Q_{l'\tau'}\over\epsilon\,
     |{\bf R}_{l\tau}-{\bf R}_{l'\tau'}|} \;,
\end{equation}
where ${\bf R}_{l\tau}$ is the position in the ideal cubic structure
of the atom on site $\tau$ of cell $l$
($\tau$=\{A,B,O$_1$,O$_2$,O$_3$\}),
and $\epsilon$ is a dielectric constant providing some screening
effects.  While we find that
this model can be successfully applied to other cases
(see below), we first limit ourselves to divalent-A
compounds having a common A atom on all A sites, and compositional
freedom on the B site. In this case, we have $Q_{l,\rm A}=q_{\rm
A}=+2$ and $Q_{l,\rm O}=q_{\rm O}=-2$ independent of $l$.  Focusing
on the B sites on which the alloying occurs, we can decompose
$Q_{l,\rm B}$ as
\begin{equation}
Q_{l,\rm B}=q_{\rm B}+\Delta q_l \;,
\end{equation}
where the {\it average} valence $q_{\rm B}$ is equal to $+4$
while $\Delta q_l$ depends on $l$.  In order to preserve overall
charge neutrality, the average $\overline{\Delta q}$ of
$\Delta q_l$ over all cells must vanish.

Substituting expression (2) for $Q_{l,\rm B}$ into Eq.~(1),
we can write $E=E_0+E_1+E_2$,
where the subscript refers to the number of powers of $\Delta q$
appearing in that term.  Then $E_0$ is just a constant; it is
the energy, within this model, of an ideal divalent-A and
tetravalent-B perovskite compound.  $E_1$ can be written as
\begin{equation}
 E_1 = {e^2\over2} \sum_l \Delta q_l
   \left[ \sum_{l'\tau'} {q_{\tau'}\over\epsilon\,
     |{\bf R}_{l\tau}-{\bf R}_{l'\tau'}|} \right] \;,
\end{equation}
but noting that the term in brackets does not actually depend on $l$
and using $\overline{\Delta q}=0$, it follows that $E_1$ vanishes.
Thus the only term depending on the atomic order is
\begin{equation}
 E_2 = {e^2\over2} \sum_{l\ne l'} {\Delta q_l \Delta q_l'\over\epsilon\,
     |{\bf R}_{l}-{\bf R}_{l'}|} \;.
\end{equation}
In fact, using ${\bf R}_l={\bf l}a$ where
$a$ is the lattice constant,
\begin{equation}
E_2 = {e^2\over2\epsilon a} \sum_{l\ne l'}
    {\Delta q_l \Delta q_l'\over |\bf l-l'|} \;.
\end{equation}
Eq.~(5) demonstrates that our model can be reduced to focus only on
the alloying sublattice (e.g., the B sublattice in the present
case).  Moreover, the model is parameter-free in the sense that
$e^2/2\epsilon a$ defines a characteristic energy that will set
the temperature scale on which ordering will occur.

In the following, we apply our model to various cases
of divalent-A compounds with B-site alloying. We thus
introduce a B-site-only notation.  For example,
IV$_x$IV$'$$_{1-x}$ will denote a homovalent binary alloy
having tetravalent B atoms, e.g., Pb(Zr$_{x}$Ti$_{1-x}$)O$_{3}$.
In the same manner, II$_{(1-x)/3}$IV$_x$V$_{2(1-x)/3}$ will
indicate a heterovalent ternary, such as
$(1-x)$~Ba(Mg$_{1/3}$Nb$_{2/3}$)O$_{3}$~+~$x$~BaZrO$_{3}$, which is
the result of mixing a II$_{1/3}$V$_{2/3}$ heterovalent
binary (Ba(Mg$_{1/3}$Nb$_{2/3}$)O$_{3}$) with a tetravalent-B
bulk (BaZrO$_{3}$).

In view of Eqs.~(2) and (5),
two points are immediately obvious.
(i) The present model does not allow IV$_x$IV$'$$_{1-x}$ homovalent
binary alloys to order, since $\Delta q_l$ is identically zero
in all cells $l$.  This is in agreement with the
experimental absence of LRO in homovalent alloys \cite{Cross}.
(ii) II$_{1/2}$VI$_{1/2}$ and III$_{1/2}$V$_{1/2}$ heterovalent
binaries should behave much the same, since $\Delta q_l$ consists
in both cases of equal populations of charge $\pm n$, where $n=2$
and 1 for II$_{1/2}$VI$_{1/2}$ and III$_{1/2}$V$_{1/2}$ alloys,
respectively.  Thus, these two cases really only differ in the
magnitude of the effective energy scale factor $n^2e^2/2\epsilon a$.
By computing the energy of a large variety of candidate structures and 
by performing Monte-Carlo simulations for small temperatures (see below), 
we determined that the ground-state (zero-temperature) structure
predicted by the present model in these two cases is rocksalt,
i.e., an alternation of $\pm n$ charged planes along the [111] direction.
This is precisely the LRO observed experimentally in
Pb(Sc$_{1/2}$Ta$_{1/2}$)O$_{3}$ \cite{Setter,Cross2},
Pb(In$_{1/2}$Nb$_{1/2}$)O$_{3}$ \cite{Cross},
Pb(Yb$_{1/2}$Ta$_{1/2}$)O$_{3}$ \cite{Yasuda},
Pb(Mg$_{1/2}$W$_{1/2}$)O$_{3}$ \cite{Cross},
and Pb(Co$_{1/2}$W$_{1/2}$)O$_{3}$ \cite{Cross},

Similar calculations show that the present model also reproduces 
the ground-state 1:2 structure observed in
many II$_{1/3}$V$_{2/3}$ heterovalent binaries, such as
Ba(Mg$_{1/3}$Nb$_{2/3}$)O$_{3}$ (BMN) \cite{Akbas2},
Ba(Zn$_{1/3}$Ta$_{2/3}$)O$_{3}$ (BZT) \cite{Jacobson,Allen},
Ba(Mg$_{1/3}$Ta$_{2/3}$)O$_{3}$ (BMT) \cite{Guo},
Ba(Sr$_{1/3}$Ta$_{2/3}$)O$_{3}$ (BST) \cite{Guo},
Ba(Zn$_{1/3}$Nb$_{2/3}$)O$_{3}$ (BZN) \cite{Treiber},
Ba(Ni$_{1/3}$Nb$_{2/3}$)O$_{3}$ (BNN) \cite{Kim},
and Ca(Ca$_{1/3}$Ta$_{2/3}$)O$_{3}$ (CCT) \cite{Burton3}.
That is, we do
indeed find that the structure minimizing the electrostatic
energy consists of one (111) layer of $-2$ relative charges
(e.g., Zn-derived in BZN) alternating with two (111) layers
of $+1$ relative charges (e.g., Nb-derived in BZN) along the
[111] axis.

To extract further consequences from the model, it is necessary to
solve for its thermodynamic behavior as a function of temperature.
This is done using Metropolis Monte Carlo simulations
\cite{Metropolis} in which the (electrostatic) energy
$E_2$ is given by Eq.~(5).  To be definite, we assign
$a$=7.7~a.u.\ and $\epsilon$=10.  We normally limit ourselves
to 6$\times$6$\times$6 supercells
since we checked that increasing the size of the calculations up to
12$\times$12$\times$12
supercells does not change the results appreciably.  We typically
use a very large number of trial moves (up to $\sim 3\times10^7$
in transition regions) in order to obtain good statistics
for each temperature. Since this corresponds to a large number
($\sim$35,000) of sweeps, the initial configuration is unimportant;
it can either be chosen at random, or set to a given atomic order
such as 1:2 or 1:1.  To find the ground state, we first perform
calculations at high temperature (e.g., 4,000K), and then very
slowly decrease the temperature until reaching  a structure for
which no moves are accepted by the Monte-Carlo program.

Once the thermodynamic state corresponding to a given
temperature has been reached,  the search for LRO is done by
calculating, from the Monte-Carlo outputs, the Fourier transform
of the charge-charge correlation function,
\begin{equation}
\eta({\bf k}) = \alpha \sum_{ll'}
\Delta q_l \, \Delta q_{l+l'}
\, \exp(-i \bf k\cdot l') \;.
\end{equation}
Here $\alpha$ is a normalization factor, the sum runs over
the charges in the B sublattice, and $\bf k$ is the wavevector
in the Brillouin zone of the unit cubic cell.
It can easily be demonstrated that $\eta({\bf k})$ is
directly proportional to the ensemble average of the square of
the Fourier transform of the charge distribution.  Thus, a
large value of $\eta$ at
${\bf k}=2\pi({1\over2},{1\over2},{1\over2})$
corresponds to strong 1:1 LRO along the [111] direction, while a
peak in $\eta$ at
${\bf k}=2\pi({1\over3},{1\over3},{1\over3})$
indicates 1:2 order along the [111] direction.

Figures 1(a-c) show the predicted behavior of $\eta({\bf k})$
vs.\ tetravalent composition $x$ at $T$=1000K for
III$_{(1-x)/2}$IV$_x$V$_{(1-x)/2}$,
II$_{(1-x)/2}$IV$_x$VI$_{(1-x)/2}$, and
II$_{(1-x)/3}$IV$_x$V$_{2(1-x)/3}$ heterovalent ternaries,
respectively.
We find that progressively increasing $x$ in
III$_{(1-x)/2}$IV$_x$V$_{(1-x)/2}$ and
II$_{(1-x)/2}$IV$_x$VI$_{(1-x)/2}$ leads to a continuous transition
from rocksalt-type order to a disordered state.  This is consistent
with the experimental findings in $(1-x)$
Pb(Sc$_{1/2}$Ta$_{1/2}$)O$_{3}$ + $x$ PbTiO$_{3}$
\cite{Burton3,Giniewicz}.
We also predict that the 1:1 rocksalt order can survive
in II$_{(1-x)/2}$IV$_x$VI$_{(1-x)/2}$ alloys up to rather
large $x$.  This compositional difference between
III$_{(1-x)/2}$IV$_x$V$_{(1-x)/2}$ and
II$_{(1-x)/2}$IV$_x$VI$_{(1-x)/2}$ ternaries is due to the fact
that the ground state energy of the latter is
four times deeper than that of the former (for a given $\epsilon$).
Thus, quite a large number of tetravalent atoms is required
to induce a disordered state in the
II$_{(1-x)/2}$IV$_x$VI$_{(1-x)/2}$ ternaries.
We varied the temperature for the III$_{(1-x)/2}$IV$_x$V$_{(1-x)/2}$
and II$_{(1-x)/2}$IV$_x$VI$_{(1-x)/2}$ ternaries but found no
qualitative changes (e.g., new phases), although naturally the
transition from the ordered to the disordered phase is shifted
to lower $x$ with increasing $T$\cite{explan-t}.

%
\begin{figure}
\epsfxsize=2.2 truein
\centerline{\epsfbox{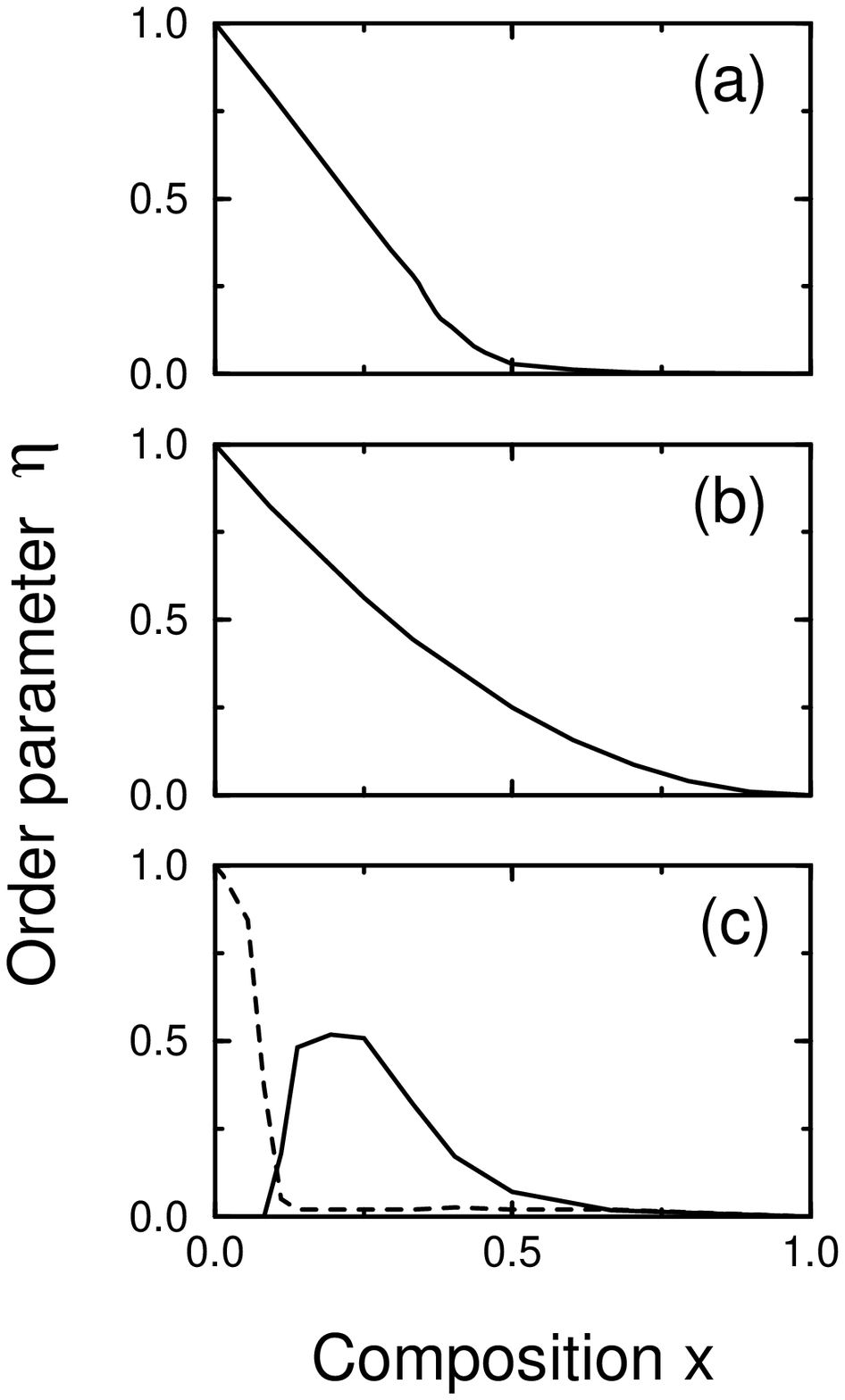}}
\vskip 0.2truein
\caption{Monte Carlo simulations of the long-range order parameter
$\eta({\bf k})$ {\it vs.} tetravalent atomic composition $x$
for the model of Eq.~(5), with 6$\times$6$\times$6 supercells,
$a$=7.7~a.u., $\epsilon$=10, and $T$=1000K.
Solid and dashed lines refer to 1:1 order
[${\bf k}=2\pi({1\over2},{1\over2},{1\over2})$] and 1:2 order
[${\bf k}=2\pi({1\over3},{1\over3},{1\over3})$], respectively.
(a) III$_{(1-x)/2}$IV$_x$V$_{(1-x)/2}$ ternaries.
(b) II$_{(1-x)/2}$IV$_x$VI$_{(1-x)/2}$ ternaries.
(c) II$_{(1-x)/3}$IV$_x$V$_{2(1-x)/3}$ ternaries.}
\label{fig1}
\end{figure}

Figure 1(c) demonstrates that the atomic ordering predicted in
II$_{(1-x)/3}$IV$_x$V$_{2(1-x)/3}$ ternaries is even richer than
in the previous two cases.  We find 1:2 order along [111]
for small $x$, then 1:1 rocksalt order for intermediate
$x$, and finally a disordered state for even larger $x$.  These
predictions are in remarkable agreement with recent experimental
observations of precisely this ordering sequence as a function
of $x$ in the microwave ceramics
$(1-x)$ Ba(Mg$_{1/3}$Nb$_{2/3}$)O$_{3}$ + $x$ BaZrO$_{3}$ \cite{Akbas2},
$(1-x)$ Ba(Mg$_{1/3}$Ta$_{2/3}$)O$_{3}$ + $x$ BaZrO$_{3}$ \cite{Chai},  and
$(1-x)$ Ba(Zn$_{1/3}$Ta$_{2/3}$)O$_{3}$ + $x$ BaZrO$_{3}$ \cite{Davies1}.
We find numerically that the critical value of $x$ at which the
crossover occurs from 1:2 to 1:1 order depends on the
temperature used in the Monte-Carlo simulation,
as well as on the value of the dielectric constant in Eq. (5).
Increasing the temperature or increasing the dielectric constant
leads to a decrease of this crossing composition.  For example,
keeping the lattice constant equal to 7.7 a.u.\ and using a
dielectric constant of 15 yields a critical composition
$x_c\sim 5\%$ for $T$=1000K.  (A further increase in temperature
or dielectric constant leads to alloy disorder even at
small $x$ composition.)  Moreover, our calculations also confirm
a suggestion based on experimental observations \cite{Akbas2,Chai}
for the structure responsible for the 1:1 order.  Specifically,
when we observe 1:1 order for $x<25\%$, we find a rocksalt
structure in which one sublattice is almost entirely
occupied by pentavalent atoms, while the second sublattice
is basically composed of the divalent and tetravalent atoms
plus the remaining pentavalent atoms.
The figure also indicates that the strongest 1:1 order occurs
for a bulk composition of 25\%, which is consistent with the
experimental findings that the 1:1-order X-ray reflections
seems to reach a maximum for $x=0.25$ in
$(1-x)$~Ba(Zn$_{1/3}$Ta$_{2/3}$)O$_{3}$~+~$x$~BaZrO$_{3}$ (BZT-BZ)
\cite{Chai}.

On the other hand, there is a noteworthy exception to 
the ordering sequence shown in Fig.~1(c).  Namely,
in II$_{(1-x)/3}$IV$_x$V$_{2(1-x)/3}$ ternaries for which the divalent
A atom is Pb, the 1:2 ordering is {\it not} observed as $x\rightarrow0$.
Instead, these compounds exhibit weak 1:1 X-ray reflections
all the way down to $x=0$, as observed in
Pb(Mg$_{1/3}$Nb$_{2/3}$)O$_{3}$
(PMN) \cite{Husson1,Husson2,Chen} and
Pb(Mg$_{1/3}$Ta$_{2/3}$)O$_{3}$
(PMT) \cite{Akbas1}.
As this appears to be at variance with our model, some 
explanation is clearly required.  First, we note that our model
predicts that the 1:2 and 1:1 ordering types are very close in free
energy in II$_{(1-x)/3}$IV$_x$V$_{2(1-x)/3}$ ternaries at small $x$.
It is thus possible that in II$_{1/3}$V$_{2/3}$ binaries for which
the A sublattice is occupied by Pb,
some physical mechanism may occur that is responsible for triggering
the 1:1 order in place of the 1:2 type.  For example,
a possible mechanism might be the covalency induced by 
the existence of short Pb--O bonds in the lead compounds \cite{Egami2}.
The difference in covalency between Pb compounds and Ba compounds
has been previously proposed to explain the difference in the
ferroelectric behavior of PbTiO$_{3}$ and BaTiO$_{3}$ \cite{Cohen2}. 

However, we wish to propose another possible mechanism for the
existence of weak 1:1 order in PMN and PMT, related to the
multivalent nature of the Pb atom.  As is well known, Pb can be
either divalent {\it or tetravalent}, in view of its $s^2p^2$
valence electronic configuration.  Indeed, as much as 5\% of
Pb$^{+4}$ has been experimentally reported in PMN \cite{Egami}.
These tetravalent Pb ions were initially thought to be sitting on the
A sublattice of the perovskite alloy \cite{Egami}.  However,
very recent experiments on Pb(Sc$_{1/2}$Ta$_{1/2}$)O$_{3}$
(PST) films indicate the presence of some Pb$^{+4}$ {\it on the B
sublattice} of the perovskite structure\cite{Whatmore}.   Thus,
we would like to raise the possibility that the Pb$^{+4}$
atoms in PMN sit on the B sublattice.  In this way, the application of
Eq.~(2) to PMN must recognize the existence of three species
on the B sublattice: Mg$^{+2}$, Nb$^{+5}$, and a small number of
Pb$^{+4}$ (with relative charges $-2$, $+1$, and 0, respectively).
The prediction of
the present model for this case is again just that indicated in Fig.~1(c):
only a small amount of Pb$^{+4}$ (i.e., a small value of $x$) is needed
to give rise to the weak 1:1 order observed
in PMN and PMT.  Interestingly, the existence of Pb$^{+4}$ on
the B sublattice of II$_{1/2}$VI$_{1/2}$ and III$_{1/2}$V$_{1/2}$
heterovalent alloys, e.g. Pb(Co$_{1/2}$W$_{1/2}$)O$_{3}$ (PCW) and
PST, will not lead to a change
of ordering type.  This is clearly shown in Figs.~1(a-b):
whether $x$ is small or exactly equal to zero, the 1:1 order
is the only ordering predicted by our model for
II$_{(1-x)/2}$IV$_x$VI$_{(1-x)/2}$ and
III$_{(1-x)/2}$IV$_x$V$_{(1-x)/2}$ ternaries.  In other words,
unlike in PMN, the inclusion of a small amount of
Pb$^{+4}$ ions in PST may change the strength of the ordering
but will not change the nature of the order itself.

While we have chosen to focus above on non-magnetic perovskite
insulators with B-site substitutional disorder, our model is by no
means limited to this case.  In fact, the substitutional
``impurity'' may even be a vacancy.  Denoting the latter by [~], our
model predicts that the ground state of
(Ca$_{1/2}$[~]$_{1/2}$)TaO$_{3}$ should adopt the 1:1 rocksalt
order of the Ca atoms and vacancies on the A sites. This is exactly
the type of LRO observed experimentally in this system
\cite{Burton3}.  The extensions to magnetic perovskite insulators
or to simultaneous A-site and B-site disorder, with possible
incorporation of spin contributions or covalent effects into the model,
are promising avenues for future work. 

In summary, we have demonstrated that a purely ionic model is
able to reproduce the various types of B-site compositional
LRO observed in a wide variety of binary and ternary perovskite
alloys.  The remarkable success of
this approach strongly suggests that the main driving mechanism
responsible for the LRO occurring in heterovalent perovskite
alloys is simply the electrostatic interaction between the
different species.  We also raise the possibility of a new
mechanism in which a small amount of Pb$^{+4}$ on the B sublattice
may be responsible for (or at least contribute to) the weak 
1:1-order X-ray reflections observed in some Pb relaxor systems.

This work is supported by the Office of Naval Research grant
N00014-97-1-0048.  We would like to acknowledge B.P.~Burton and
T.~Egami for many very useful discussions.



\begin{references}
\bibitem{explan-t} In fact, Figs.~1(a) and (b) can be regarded as
giving the results for the
II$_{(1-x)/2}$IV$_x$VI$_{(1-x)/2}$ and
III$_{(1-x)/2}$IV$_x$V$_{(1-x)/2}$
ternaries at $T$=4000K and $T$=250K respectively.
\end{references}
\end{document}